\documentclass[preprint,12pt]{article}
\usepackage{amssymb,amsmath,amsthm,amsfonts,amscd}
\usepackage{graphicx,times}
\newcommand\be{\begin{equation}}
\newcommand\ee{\end{equation}}
\newcommand\bea{\begin{eqnarray}}
\newcommand\eea{\end{eqnarray}}

\newcommand{\fatalpha}{{\bf \alpha \kern -0.44em \alpha}}
\newcommand{\fatsigma}{{\bf \sigma \kern -0.54em \sigma}}
\newcommand{\tpchi}{{\bf D \kern -0.35em D}}
\newcommand{\llambda}{{\bf \lambda \kern -0.45em \lambda}}

\bibliography{plain}
\title{\bf Entanglement Witnesses and Characterizing Entanglement Properties of Some PPT States}\vspace{20mm}
\author{ M. A. Jafarizadeh $^{a,b,c}$
 \thanks{E-mail:jafarizadeh@tabrizu.ac.ir}  ,   N. Behzadi$^{a,b}$  \thanks{E-mail:behzadi@tabrizu.ac.ir}  ,  Y. Akbari$^{d}$  \thanks{E-mail:yakbari@azaruniv.edu}
 \\ $^a${\small Department of Theoretical Physics and Astrophysics,
Tabriz University, Tabriz 51664, Iran.} \\ $^b${\small Institute
for Studies in Theoretical Physics and Mathematics, Tehran
19395-1795, Iran.} \\ $^c${\small Research Institute for
Fundamental Sciences, Tabriz 51664, Iran.}  \\ $^d${\small
Department of Physics, Azarbaijan University of Tarbiat Moallem,
53714-161, Tabriz, Iran.}} \pagebreak

\vspace{20mm}
\begin{document}
\maketitle \vspace{15mm}
\newpage
\begin{abstract}
On the basis of linear programming, new sets of entanglement witnesses (EWs) for $3\otimes3$ and $4\otimes4$ systems are constructed. In both cases, the constructed EWs correspond to the hyper-planes contacting, without intersecting, the related feasible regions at line segments and restricted planes respectively. Due to the special property of the contacting area between the hyper-planes and the feasible regions, the corresponding hyper-planes can be turned around the contacting area throughout a bounded interval and hence create an infinite number of EWs. As these EWs are able to detect entanglement of some PPT states, they are non-decomposable (nd-EWs).
\end{abstract}



\vspace{70mm}
\newpage
\section{Introduction}
Entanglement which appears only in composite quantum systems, is the main difference between quantum mechanics and classical physics. It is used as a physical resource to realize various quantum information and quantum computation tasks such as quantum cryptography, teleportation, dense coding, and key distribution \cite{nielsen1,ekert1,preskill}. The fundamental problem of entanglement theory is the finding out separating boundary of the entangled states and  separable ones. The celebrated Peres-Horodecki criterion based on positive partial transpose (PPT) determines this boundary for $2\otimes2$ and $2\otimes3$ cases \cite{peres,phorodecki} but it has no efficiency for PPT entangled states appearing in the higher dimensional systems. On the other hand, distinguishing between the PPT entangled states which can not be distilled and those which can be distilled is an important problem in quantum communication field \cite{mprhorodecki1}. The most general approach to study the entanglement of quantum states in higher dimensional physical systems is based on the notion of entanglement witnesses (EWs) \cite{mprhorodecki,terhal}. The EWs are essential tools in entanglement theory since it has been shown that for any entangled state there exists at least one EW which detects its entanglement \cite{mprhorodecki,woronowicz}. A Hermitian operator W is said to be an EW if and only if for all separable states $\rho_{sep}$, $Tr(W\rho_{sep})\geq0$ and at least for one
entangled state $\rho_{ent}$, $Tr(W\rho_{ent})<0$ ( one says that $\rho_{ent}$ is detected by W). Clearly, the construction of EWs is a hard task. Although it is easy to construct a Hermitian operator W which has negative expectation value with some entangled states but it is very difficult to check that its expectation values with all separable states are non-negative. So, several approaches for constructing EWs have been proposed. It has been shown that among these approaches, the linear programming, as a special case of convex optimization \cite{boyd,chong}, is a very useful one \cite{jafari1,jafari2,jafari3,jafari4}. In linear programming method, the main problem is the determination of boundaries of a convex set known as feasible region. The feasible region is associated to the convex set of separable states. The boundaries of the feasible region are characterized by the operators used in constructing the EWs. When the feasible region was specified then every hyperplane tangent to the feasible region corresponds to an EW. A hyperplane may be contact the feasible region at a point, line segment, restricted plane, etc. If the boundary is differentiable on the contact area, the hyper-plane has to be fixed, otherwise it can be turned around the contact area throughout a bounded interval. The EWs constructed so far via linear programming, correspond to the fixed tangent hyper-planes. When a boundary of the fesible region is a hyper-plane by itself, it is called an exact boundary.
\par
In this paper, by choosing suitable operators, new sets of EWs are constructed for $3\otimes3$ analytically and for $4\otimes4$ numerically. These EWs correspond to the hyper-planes which can turned around the contact line segments or restricted planes for the $3\otimes3$ and $4\otimes4$ cases respectively. It is shown that these hyper-planes can be turned around in a bounded interval such that they do not intersect the feasible region. In this way, we obtain an infinite number of hyper-planes which are in contacting with the feasible region and hence an infinite number of EWs for $3\otimes3$ and $4\otimes4$ systems. Finally, we show that these EWs are nd-EWs since they are able to detect the entanglement of some PPT sates.
\par
The paper is organized as follow: In section $2$, for $3\otimes3$ case, we obtain a part of the boundary of the feasible region as a exact boundary. The intersection of this boundary by the other boundaries is three line segments. We see that the planes which contact the feasible region only at these line segments, correspond to new EWs. Section $3$ is devoted to extend the approach of section $2$, to the $4\otimes4$ systems. It is ended by a brief conclusion.
\section{Entanglement Witnesses For $3\otimes3$ Systems}
In this section we are going to introduce a method based on linear programming to construct new EWs on $3\otimes3$ Hilbert space which are able to detect the entanglement of some PPT states. When one deals with linear programming, the aim is optimization of a linear function under some linear constraints. The linear constraints define a region called feasible region. So to reduce the construction of EWs to a linear programming problem, the related feasible region must be characterized. As mentioned in introduction, the shape of the boundaries of the feasible region depends on the operators chosen for constructing of EWs. So in this paper, we choose the following operators
\begin{equation}\label{eq1}
\begin{array}{c}
  O_{1}=\frac{1}{3}(|01\rangle\langle01|+|12\rangle\langle12|+|20\rangle\langle20|) \\
  O_{2}=\frac{1}{3}(|02\rangle\langle02|+|10\rangle\langle10|+|21\rangle\langle21|) \\
  O_{3}=|\psi\rangle\langle\psi| \\
\end{array}
\end{equation},
where
$$
|\psi\rangle=\frac{1}{\sqrt{3}}(|00\rangle+|11\rangle+|22\rangle).
$$
It is seen that these operators are orthogonal to each other, i.e. $O_{i}O_{j}=0$, $i\neq j=1,2,3$, and
$O_{2}=\Pi O_{1}\Pi$, $O_{3}=\Pi O_{3}\Pi$ where $\Pi$ is an operator which permutes the particles. We denote the expectation values of the above operators with pure product states $|\alpha\rangle\otimes|\beta\rangle$ by $p_{1}$, $p_{2}$, $p_{3}$ respectively as
\begin{equation}\label{eq2}
\begin{array}{c}
  p_{1}=\frac{1}{3}(|\alpha_{0}|^{2}|\beta_{1}|^{2}+|\alpha_{1}|^{2}|\beta_{2}|^{2}+|\alpha_{2}|^{2}|\beta_{0}|^{2}) \\
  p_{2}=\frac{1}{3}(|\alpha_{0}|^{2}|\beta_{2}|^{2}+|\alpha_{1}|^{2}|\beta_{0}|^{2}+|\alpha_{2}|^{2}|\beta_{1}|^{2}) \\
  p_{3}=\frac{1}{3}|\alpha_{0}\beta_{0}+\alpha_{1}\beta_{1}+\alpha_{2}\beta_{2}|^{2}.
\end{array}
\end{equation}
The $p_{1}$, $p_{2}$ and $p_{3}$ can be considered as components of a point lying in the three dimensional Euclidean space. Hence, the set of all pure product states form a region in this Euclidean space. Since every separable state can be written as a convex combination of pure product states, the feasible region is the convex hull of the mentioned region. Now let us go to figure out a part of the boundary of the feasible region which has essential importance in constructing our EWs. As the maximum value of $p_{1}$, $p_{2}$ and $p_{3}$ is $\frac{1}{3}$, the points ($\frac{1}{3},0,0$), ($0,\frac{1}{3},0$) and ($0,0,\frac{1}{3}$) which correspond to the product states listed in the table $1$, are vertices of the feasible region. The plane which passes through these points has the following equation
\begin{equation}\label{eq3}
3(p_{1}+p_{2}+p_{3})=1
\end{equation}
To find out that this plane is a boundary of the feasible region, the linear function on the left hand side of the (\ref{eq3}) must be maximized with respect to the product states. If the maximum value is one then the plane is a part of the boundary of the feasible region, otherwise it is not. Calculation shows that the maximum value is greater than one and hence this plane is not a boundary for the feasible region. The product which gives the maximum value is
$$
\begin{array}{c}
  \frac{1}{\sqrt{3}}\left(%
\begin{array}{c}
  1 \\
  1 \\
  1 \\
\end{array}%
\right)\otimes\frac{1}{\sqrt{3}}\left(%
\begin{array}{c}
  1 \\
  1 \\
  1 \\
\end{array}%
\right) \\
\end{array}
$$
and its corresponding point is ($\frac{1}{9},\frac{1}{9},\frac{1}{3}$). This point is another vertex of the feasible region which also given in the table $1$. The plane tangent to the feasible region at this point is
\begin{equation}\label{eq4}
3(p_{1}+p_{2}+p_{3})=\frac{5}{3}
\end{equation}
This means that for all separable states, we have always $3(p_{1}+p_{2}+p_{3})\leq\frac{5}{3}$. In the next step, we note that the equation of the plane passing through the vertices ($\frac{1}{3},0,0$), ($0,\frac{1}{3},0$) and  ($\frac{1}{9},\frac{1}{9},\frac{1}{3}$) is
\begin{equation}\label{eq5}
3(p_{1}+p_{2})+p_{3}=1
\end{equation}
The maximum value of the left hand side of this equation with respect to the product states is one so this plane is an exact boundary of the feasible region and any separable state always satisfies the inequality $3(p_{1}+p_{2})+p_{3}\leq1$. The existence of this exact plane makes a good possibility for constructing a new type of EWs.  The line segments surrounding the plane (\ref{eq5}), are also on the boundary of the feasible region. The sets of planes which are in contact with the feasible region at the mentioned line segments and not intersect the feasible region can be corresponded to new type of EWs which we desire to construct them. To this aim, we proceed as follows. First, we choose the vertices ($0,\frac{1}{3},0$), ($0,0,\frac{1}{3}$) and ($\frac{1}{9},\frac{1}{9},\frac{1}{3}$). The plane passing through them is
\begin{equation}\label{eq6}
3(-p_{1}+p_{2}+p_{3})=1
\end{equation}
Maximizing the left hand side of this equation with respect to the product states, we obtain the maximum value $\frac{5}{4}$ at the following state
$$
\begin{array}{c}
  \frac{1}{\sqrt{2}}\left(%
\begin{array}{c}
  0 \\
  \sqrt{3} \\
  1 \\
\end{array}%
\right)\otimes\frac{1}{\sqrt{2}}\left(%
\begin{array}{c}
  0 \\
  1 \\
  \sqrt{3} \\
\end{array}%
\right) \\
\end{array}
$$
So the plane (\ref{eq6}) is not a boundary plane and the corresponding point of the above pure product state, i.e. ($\frac{1}{48},\frac{3}{16},\frac{1}{4}$), is a vertex of the feasible region. The plane tangent to the feasible region at this point is
\begin{equation}\label{eq7}
3(-p_{1}+p_{2}+p_{3})=\frac{5}{4}
\end{equation}
In the second step, we choose the vertices ($0,\frac{1}{3},0$), ($\frac{1}{9},\frac{1}{9},\frac{1}{3}$) and ($\frac{1}{48},\frac{3}{16},\frac{1}{4}$). The plane passing through them is
\begin{equation}\label{eq8}
3p_{1}+9p_{2}+5p_{3}=3
\end{equation}
which it is not a boundary of the feasible region because its maximum value with respect to the product states is not $3$ but $\frac{73}{24}$. This maximum value is achieved at the point ($\frac{1}{192},\frac{49}{192},\frac{7}{48}$)  corresponding to the following product state
$$
\begin{array}{c}
  \frac{1}{4}\left(%
\begin{array}{c}
  0 \\
  \sqrt{2} \\
  \sqrt{14} \\
\end{array}%
\right)\otimes\frac{1}{4}\left(%
\begin{array}{c}
  0 \\
  \sqrt{14} \\
  \sqrt{2} \\
\end{array}%
\right) \\
\end{array}
$$
The plane tangent to the feasible region at this point is
\begin{equation}\label{eq9}
3p_{1}+9p_{2}+5p_{3}=\frac{73}{24}
\end{equation}
Finally, we choose the vertices ($0,0,\frac{1}{3}$), ($\frac{1}{9},\frac{1}{9},\frac{1}{3}$) and ($\frac{1}{48},\frac{3}{16},\frac{1}{4}$) and pass a plane through them. The plane is
\begin{equation}\label{eq10}
3(-p_{1}+p_{2})+6p_{3}=2
\end{equation}
Maximizing with respect to the product states shows that this plane again is not a boundary plane because the point ($\frac{3}{64},\frac{25}{192},\frac{5}{16}$) corresponding to the product state
$$
\begin{array}{c}
  \frac{1}{4}\left(%
\begin{array}{c}
  0 \\
  \sqrt{6} \\
  \sqrt{10} \\
\end{array}%
\right)\otimes\frac{1}{4}\left(%
\begin{array}{c}
  0 \\
  \sqrt{10} \\
  \sqrt{6} \\
\end{array}%
\right) \\
\end{array}
$$
, gives the value $\frac{17}{8}$. Hence, the tangent plane at the point ($\frac{3}{64},\frac{25}{192},\frac{5}{16}$) is
\begin{equation}\label{eq11}
3(-p_{1}+p_{2})+6p_{3}=\frac{17}{8}
\end{equation}
Now we find the intersection of the planes (\ref{eq9}), (\ref{eq11}) and the plane $p_{3}=\frac{1}{3}$ which is the vertex ($\frac{1}{12},\frac{1}{8},\frac{1}{3}$). Clearly this vertex does not belong to the feasible region as it can not be obtained from any product state. Therefore, the equation of the plane passing through the vertices ($\frac{1}{9},\frac{1}{9},\frac{1}{3}$), ($0,\frac{1}{3},0$) and ($\frac{1}{12},\frac{1}{8},\frac{1}{3}$) is
\begin{equation}\label{eq12}
3p_{1}+6p_{2}+3p_{3}=2
\end{equation}
 This plane has intersection with the plane (\ref{eq5}) throughout the line segment passing through the vertices ($\frac{1}{9},\frac{1}{9},\frac{1}{3}$) and ($0,\frac{1}{3},0$). In fact, the mentioned plane is tangent to the feasible region at the above line segment. Obviously, the plane (\ref{eq5}) which is an exact boundary plane of the feasible region, can be rotated around the line segment and coincide to the plane (\ref{eq12}). However, the plane (\ref{eq12}) is the final limit for the rotation of the exact plane (\ref{eq5}) since if we rotate it much more, the resultant plane intersects the feasible region. The permutation of particles changes the plane (\ref{eq12}) into the plane
\begin{equation}\label{eq13}
3p_{2}+6p_{1}+3p_{3}=2.
\end{equation}
 This plane which passes through the vertices ($\frac{1}{9},\frac{1}{9},\frac{1}{3}$), ($\frac{1}{3},0,0$) and ($\frac{1}{8},\frac{1}{12},\frac{1}{3}$), is tangent to the feasible region at the line segment passing through the vertices ($\frac{1}{9},\frac{1}{9},\frac{1}{3}$) and ($\frac{1}{3},0,0$). It is clear that the plane (\ref{eq13}) can also be obtained by rotating the exact plane (\ref{eq5}) around this line segment. By the same argument sketched for the plane (\ref{eq12}), the rotation is bounded by the plane (\ref{eq13}). To summarize, the planes obtained by rotating the exact plane around the line segment passing through the vertices ($\frac{1}{9},\frac{1}{9},\frac{1}{3}$) and ($0,\frac{1}{3},0$) can be written in the following parametric form
\begin{equation}\label{eq14}
\frac{p_{1}}{\alpha}+3p_{2}+(2-\frac{1}{3\alpha})p_{3}=1,
\end{equation}
where $\alpha$ varies in the interval [$\frac{1}{3} , \frac{2}{3}$] with $\alpha=\frac{1}{3}$ for the exact plane (\ref{eq5}) and $\alpha=\frac{2}{3}$ for the plane (\ref{eq12}). Under the permutation of particles, the parametric plane (\ref{eq14}) is transformed to the plane
\begin{equation}\label{eq15}
\frac{p_{2}}{\alpha}+3p_{1}+(2-\frac{1}{3\alpha})p_{3}=1,
\end{equation}
which represents a set of planes obtained by rotating the exact plane (\ref{eq5}) around the line segment passing through the vertices ($\frac{1}{9},\frac{1}{9},\frac{1}{3}$) and ($\frac{1}{3},0,0$). Consequently, any separable state satisfies the inequalities $\frac{p_{1}}{\alpha}+3p_{2}+(2-\frac{1}{3\alpha})p_{3}\leq1$ and $\frac{p_{2}}{\alpha}+3p_{1}+(2-\frac{1}{3\alpha})p_{3}\leq1$ for $\alpha\in[\frac{1}{3} , \frac{2}{3}]$. Finally, consider the parametric plane
\begin{equation}\label{eq16}
3(p_{1}+p_{2})+\frac{1}{\alpha}p_{3}=1,
\end{equation}
for all $\alpha\in(-\infty , 0)\bigcup[1 , \infty)$ and the plane $p_{3}=0$ for $\alpha=0$. This parametric plane is representative for a set of planes which do not intersect the feasible region and obtained by rotating the exact plane (\ref{eq5}) around the line segment passing through the vertices ($\frac{1}{3},0,0$) and ($0,\frac{1}{3},0$). Therefore for any separable state, we have $3(p_{1}+p_{2})+\frac{1}{\alpha}p_{3}\leq1$ for $\alpha\neq0$ and $p_{3}\geq0$ for $\alpha=0$.
Now the description of the feasible region for our purposes is complete and hence we are going to introduce the EWs corresponding to the planes discussed above. To this aim, let us consider the following Hermitian operator
\begin{equation}\label{eq17}
W_{\alpha}=I_{3}\otimes I_{3}-\frac{1}{\alpha}O_{1}-3O_{2}-(2-\frac{1}{3\alpha})O_{3}
\end{equation}
which is correspond to the parametric plane (\ref{eq14}). By construction, the expectation value of this operator with respect to the all separable states is positive hence it can be an EW for $\alpha\in[\frac{1}{3} , \frac{2}{3}]$. The permutation of particles transforms this EW to the following one
\begin{equation}\label{eq18}
W^{'}_{\alpha}=I_{3}\otimes I_{3}-\frac{1}{\alpha}O_{2}-3O_{1}-(2-\frac{1}{3\alpha})O_{3}
\end{equation}
which is correspond to the parametric plane (\ref{eq15}). So it has positive expectation value with respect to the all separable states therefore it can also be an EW. Finally, consider the Hermitian operator
\begin{equation}\label{eq19}
W^{''}_{\alpha}=I_{3}\otimes I_{3}-3(O_{1}+O_{2})-\frac{1}{\alpha}O_{3}
\end{equation}
with $\alpha\in(-\infty , 0)\bigcup[1 , \infty)$ corresponding to the parametric plane (\ref{eq16}) and the operator $W^{''}_{0}=O_{3}$ corresponding to the plane $p_{3}=0$. The operators $W^{''}_{\alpha}$ and $W^{''}_{0}$ are both positive so they can not serve as EWs. For $\alpha=\frac{1}{3}$, $W_{\alpha}=W^{'}_{\alpha}$ is a positive operator and we claim that for $\alpha\in(\frac{1}{3} , \frac{2}{3}]$ the EWs $W_{\alpha}$ and $W^{'}_{\alpha}$ are nd-EWs. For this purpose, let us consider the following state
\begin{equation}\label{eq20}
\rho=a_{1}O_{1}+a_{2}O_{2}+a_{3}O_{3}
\end{equation}
where $a_{i}\geq0$ for $i=1,2,3$ and $a_{1}+a_{2}+a_{3}=1$. It is easy to see that when $a_{1}a_{2}\geq a^{2}_{3}$, then $\rho$ is PPT. The expectation values of $W_{\alpha}$ and $W^{'}_{\alpha}$ with this state are
\begin{equation}\label{eq21}
Tr(W_{\alpha}\rho)=(a_{1}-a_{3})(1-\frac{1}{3\alpha})
\end{equation}
and
\begin{equation}\label{eq22}
Tr(W^{'}_{\alpha}\rho)=(a_{2}-a_{3})(1-\frac{1}{3\alpha})
\end{equation}
So the entanglement of the PPT state (\ref{eq20}) is detected by the EWs $W_{\alpha}$ for $0<a_{1}<a_{3}<\frac{1}{3}$ and $\frac{1}{3}<a_{2}<1$. It is also detected by the EWs $W^{'}_{\alpha}$ for $0<a_{2}<a_{3}<\frac{1}{3}$ and $\frac{1}{3}<a_{1}<1$. As a special case of the state (\ref{eq20}) we take the following state which was introduced in \cite{mprhorodecki2} \begin{equation}\label{eq23}
\rho_{_{\beta}}=\frac{\beta}{7}O_{1}+\frac{5-\beta}{7}O_{2}+\frac{2}{7}O_{3}
\end{equation}
with $0\leq\beta\leq5$. This state is separable for $2\leq\beta\leq3$ and not PPT for $\beta>4$ and $\beta<1$.
So the equation (\ref{eq21}) becomes
\begin{equation}\label{eq24}
Tr(W_{\alpha}\rho_{_{\beta}})=(\beta-2)(\frac{1}{7}-\frac{1}{21\alpha})
\end{equation}
The $W_{\alpha}$ for all $\alpha\in(\frac{1}{3} , \frac{2}{3}]$ detects the entanglement of $\rho_{_{\beta}}$ for $\beta<2$. So the $\rho_{_{\beta}}$ is PPT entangled for $1\leq\beta<2$ and free entangled for $0\leq\beta<1$.
On the other hand, the equation(\ref{eq22}) becomes
\begin{equation}\label{eq25}
Tr(W^{'}_{\alpha}\rho_{_{\beta}})=(3-\beta)(\frac{1}{7}-\frac{1}{21\alpha})
\end{equation}
It is also seen that $W^{'}_{\alpha}$ for all $\alpha\in(\frac{1}{3} , \frac{2}{3}]$ detects the entanglement of $\rho_{_{\beta}}$ for $\beta>3$. So the $\rho_{_{\beta}}$ is PPT entangled for $3<\beta\leq4$ and free entangled for $4<\beta\leq5$. Therefore $W_{\alpha}$ and $W^{'}_{\alpha}$ are nd-EWs for all  $\alpha\in(\frac{1}{3} , \frac{2}{3}]$ and our claim is proven. To make a comparison among the EWs $W_{\alpha}$ for $\alpha\in(\frac{1}{3} , \frac{2}{3}]$, it is seen from the equation (\ref{eq24}) that the EW $W_{\frac{2}{3}}$, i.e.
\begin{equation}\label{eq27}
W_\frac{2}{3}=I_{3}\otimes I_{3}-\frac{3}{2}O_{1}-3O_{2}-\frac{3}{2}O_{3},
\end{equation}
has the best detection of entanglement of $\rho_{\beta}$. The similar comparison can be made among the EWs $W^{'}_{\alpha}$ for $\alpha\in(\frac{1}{3} , \frac{2}{3}]$. The result is that the EW $W^{'}_{\frac{2}{3}}$, i.e.
\begin{equation}\label{eq28}
W^{'}_\frac{2}{3}=I_{3}\otimes I_{3}-\frac{3}{2}O_{2}-3O_{1}-\frac{3}{2}O_{3}
\end{equation}
is the best one. This EW is the same one that obtained numerically
by Doherty and et.al. in \cite{doherty}. To measure the EWs
$W_{\alpha}$ and $W^{'}_{\alpha}$ experimentally, they should be,
in fact, decomposed into a sum of locally measurable operators
\cite{gu1,gu2}. There exist the following decompositions for EWs
$W_{\alpha}$ and $W^{'}_{\alpha}$ as
$$W_{\alpha}=\frac{12\alpha-2}{27\alpha}I_{3}\otimes
I_{3}+\frac{6\alpha-1}{18\alpha}(\lambda_{2}\otimes\lambda_{2}-\lambda_{1}\otimes\lambda_{1}+\lambda_{5}\otimes\lambda_{5}-\lambda_{4}\otimes\lambda_{4}+\lambda_{7}\otimes\lambda_{7}-\lambda_{6}\otimes\lambda_{6})$$
\be\label{eq29}-\frac{3\alpha-5}{36\alpha}(\lambda_{3}\otimes\lambda_{3}+\lambda_{8}\otimes\lambda_{8})+\frac{\sqrt{3}(3\alpha-1)}{12\alpha}(\lambda_{3}\otimes\lambda_{8}-\lambda_{8}\otimes\lambda_{3})\ee
and
$$W_{\alpha'}=\frac{12\alpha'-2}{27\alpha'}I_{3}\otimes I_{3}+\frac{6\alpha'-1}{18\alpha'}(\lambda_{2}\otimes\lambda_{2}-\lambda_{1}\otimes\lambda_{1}+\lambda_{5}\otimes\lambda_{5}-\lambda_{4}\otimes\lambda_{4}+\lambda_{7}\otimes\lambda_{7}-\lambda_{6}\otimes\lambda_{6})$$
\be\label{eq30}-\frac{3\alpha'-5}{36\alpha'}(\lambda_{3}\otimes\lambda_{3}+\lambda_{8}\otimes\lambda_{8})-\frac{\sqrt{3}(3\alpha'-1)}{12\alpha'}(\lambda_{3}\otimes\lambda_{8}-\lambda_{8}\otimes\lambda_{3}),\ee
where $\lambda_{i}$s, $i=1,...,8$, are the basis for the $su(3)$ Lie algebra \cite{su(n)}. Therefore each EW requires ten measurement settings which should be measured locally by Alice and Bob simultaneously.
\section{Entanglement Witnesses For $4\otimes4$ Systems}
In this section we extend the approach of the previous section for $3\otimes3$ systems to the $4\otimes4$ ones. To this aim, we introduce four operators as
\begin{equation}\label{eq29}
\begin{array}{c}
O_{1}=\frac{1}{4}(|01\rangle\langle01|+|12\rangle\langle12|+|23\rangle\langle23|+|30\rangle\langle30|) \\ \\
O_{2}=\frac{1}{4}(|02\rangle\langle02|+|13\rangle\langle13|+|20\rangle\langle20|+|31\rangle\langle31|) \\ \\
O_{3}=\frac{1}{4}(|03\rangle\langle03|+|10\rangle\langle10|+|21\rangle\langle21|+|32\rangle\langle32|) \\ \\
O_{4}=|\psi\rangle\langle\psi|
\end{array}
\end{equation},
where
$$
|\psi\rangle=\frac{1}{2}(|00\rangle+|11\rangle+|22\rangle+|33\rangle)
$$
It is clear that $O_{3}=\Pi O_{1}\Pi$, $O_{2}=\Pi O_{2}\Pi$, $O_{4}=\Pi O_{4}\Pi$ and $O_{i}O_{j}=0$, $i\neq j=1,2,3,4$ where $\Pi$ is the permutation operator which permute the particles. The expectation values of these operators with respect to the product state $|\alpha\rangle\otimes|\beta\rangle$ are shown by $p_{1}$, $p_{2}$, $p_{3}$ and $p_{4}$ respectively as
\begin{equation}\label{eq30}
\begin{array}{c}
  p_{1}=\frac{1}{4}(|\alpha_{0}|^{2}|\beta_{1}|^{2}+|\alpha_{1}|^{2}|\beta_{2}|^{2}+|\alpha_{2}|^{2}|\beta_{3}|^{2}+|\alpha_{3}|^{2}|\beta_{0}|^{2}) \\
  p_{2}=\frac{1}{4}(|\alpha_{0}|^{2}|\beta_{2}|^{2}+|\alpha_{1}|^{2}|\beta_{3}|^{2}+|\alpha_{2}|^{2}|\beta_{0}|^{2}+|\alpha_{3}|^{2}|\beta_{1}|^{2}) \\
  p_{3}=\frac{1}{4}(|\alpha_{0}|^{2}|\beta_{3}|^{2}+|\alpha_{1}|^{2}|\beta_{0}|^{2}+|\alpha_{2}|^{2}|\beta_{1}|^{2}+|\alpha_{3}|^{2}|\beta_{2}|^{2}) \\
  p_{4}=\frac{1}{4}|\alpha_{0}\beta_{0}+\alpha_{1}\beta_{1}+\alpha_{2}\beta_{2}+\alpha_{3}\beta_{3}|^{2}.
\end{array}
\end{equation}
As before, the $p_{1}$, $p_{2}$, $p_{3}$ and $p_{4}$ can be considered as the components of a point in the feasible region lying in the four dimensional Euclidean space. Again, we determine a part of the boundary of the feasible region which is suitable for our purposes. By considering the points ($\frac{1}{4},0,0$), ($0,\frac{1}{4},0,0$), ($0,0,\frac{1}{4},0$) and ($0,0,0,\frac{1}{4}$) as the vertices of the feasible region, the equation of the hyperplane passing through them is
\begin{equation}\label{eq31}
4(p_{1}+p_{2}+p_{3}+p_{4})=1
\end{equation}
This hyperplane is not a part of the boundary of the feasible region because maximizing the left hand side of the above equation with respect to the product states shows that the point ($\frac{1}{16},\frac{1}{16},\frac{1}{16},\frac{1}{4}$) corresponding to the product state
$$
\begin{array}{c}
  \frac{1}{2}\left(%
\begin{array}{c}
  1 \\
  1 \\
  1 \\
  1 \\
\end{array}%
\right)\otimes\frac{1}{2}\left(%
\begin{array}{c}
  1 \\
  1 \\
  1 \\
  1 \\
\end{array}%
\right) \\
\end{array}
$$
does not lie on the hyperplane. So the mentioned point is another vertex of the feasible region. Next, if we take the vertices ($\frac{1}{4},0,0,0$), ($0,\frac{1}{4},0,0$), ($0,0,\frac{1}{4},0$) and ($\frac{1}{16},\frac{1}{16},\frac{1}{16},\frac{1}{4}$), the equation of the hyperplane passing through them is \begin{equation}\label{eq32}
4(p_{1}+p_{2}+p_{3})+p_{4}=1.
\end{equation}
This is an exact boundary hyperplane of the feasible region. Therefore, any separable state must satisfy $4(p_{1}+p_{2}+p_{3})+p_{4}\leq1$. The existence of such hyperplane as a part of the boundary of the feasible region allows us to use the approach of the previous section to construct EWs. However, due to the lose of intuition in the four dimensional case, we can not deal with this problem analytically as well as in the $3\otimes3$ case and we have to invoke to the numerical evaluation. Let us consider the three vertices ($0,\frac{1}{4},0,0$), ($0,0,\frac{1}{4},0$) and ($\frac{1}{16},\frac{1}{16},\frac{1}{16},\frac{1}{4}$) lying on the hyperplane (\ref{eq32}) and the point ($\alpha$,0,0,0) with $\alpha>\frac{1}{4}$. The hyperplane passing through these points is
\begin{equation}\label{eq33}
\frac{p_{1}}{\alpha}+4(p_{2}+p_{3})+(2-\frac{1}{4\alpha})p_{4}=1.
\end{equation}
Numerical evaluation shows that when $\frac{1}{4}<\alpha\leq\frac{1}{3}$, the hyperplane remains in contact, without intersecting, with the feasible region at the restricted plane passing through the vertices ($0,\frac{1}{4},0,0$), ($0,0,\frac{1}{4},0$) and ($\frac{1}{16},\frac{1}{16},\frac{1}{16},\frac{1}{4}$). Thus, the inequality $\frac{p_{1}}{\alpha}+4(p_{2}+p_{3})+(2-\frac{1}{4\alpha})p_{4}\leq1$ must be satisfied for any separable state. For $\alpha>\frac{1}{3}$, the mentioned hyperplane does not remain in contact with the feasible region because it intersects the feasible region. In comparison with the $3\otimes3$ case, this situation, i.e. the variation of $\alpha$ in the mentioned range, is similar to the rotation of the plane (\ref{eq14}) around the line segment which was discussed in the previous section. Permutation of the particles transforms the hyperplane (\ref{eq33}) into the following one
\begin{equation}\label{eq34}
\frac{p_{3}}{\alpha}+4(p_{1}+p_{2})+(2-\frac{1}{4\alpha})p_{4}=1.
\end{equation}
In this case, numerical evaluation shows that, as for the hyperplane (\ref{eq33}), when $\frac{1}{4}<\alpha\leq\frac{1}{3}$ the hyperplane remains in contact with the feasible region at the restricted plane passing through the vertices ($\frac{1}{4},0,0,0$), ($0,\frac{1}{4},0,0$) and ($\frac{1}{16},\frac{1}{16},\frac{1}{16},\frac{1}{4}$). Hence, for any separable state we have $\frac{p_{3}}{\alpha}+4(p_{1}+p_{2})+(2-\frac{1}{4\alpha})p_{4}\leq1$. For $\alpha>\frac{1}{3}$, the hyperplane intersects the feasible region. If we take the vertices ($\frac{1}{4},0,0,0$), ($0,0,\frac{1}{4},0$) and ($\frac{1}{16},\frac{1}{16},\frac{1}{16},\frac{1}{4}$) and the point ($0,\alpha,0,0$) with $\alpha>\frac{1}{4}$, the passing hyperplane through them is
\begin{equation}\label{eq35}
\frac{p_{2}}{\alpha}+4(p_{1}+p_{3})+(2-\frac{1}{4\alpha})p_{4}=1
\end{equation}
This hyperplane is invariant under the permutation of particles. Numerical results show that for all $\alpha\geq\frac{1}{4}$, the hyperplane in (\ref{eq35}) remains tangent to the feasible region, hence any separable state satisfies the inequality $\frac{p_{2}}{\alpha}+4(p_{1}+p_{3})+(2-\frac{1}{4\alpha})p_{4}\leq1$. The contact area is the restricted plane passing through the mentioned three vertices. Finally, the hyperplane passing through the other vertices ($\frac{1}{4},0,0,0$), ($0,\frac{1}{4},0,0$), ($0,0,\frac{1}{4},0$) and the point ($0,0,0,\alpha$) is
\begin{equation}\label{eq36}
4(p_{1}+p_{2}+p_{3})+\frac{p_{4}}{\alpha}=1
\end{equation}
where $\alpha\in(-\infty , 0)\bigcup[1 , \infty)$. For $\alpha=0$, we have the hyper-plane $p_{4}=0$ as a part of the boundary of the feasible region because every separable state satisfies the $p_{4}\geq0$. When $\alpha$ varies in this range, the corresponding hyper-planes remain in contact with the feasible region at the restricted plane passing through the vertices ($\frac{1}{4},0,0,0$), ($0,\frac{1}{4},0,0$) and ($0,0,\frac{1}{4},0$). Now we are ready to construct our EWs. The EW  corresponding to the hyperplane (\ref{eq33}) is
\begin{equation}\label{eq37}
W_{\alpha}=I_{4}\otimes I_{4}-\frac{1}{\alpha}O_{1}-4(O_{2}+O_{3})-(2-\frac{1}{4\alpha})O_{4},
\end{equation}
Permutating of the particles gives the EW
\begin{equation}\label{eq38}
W^{'}_{\alpha}=I_{4}\otimes I_{4}-\frac{1}{\alpha}O_{3}-4(O_{1}+O_{2})-(2-\frac{1}{4\alpha})O_{4},
\end{equation}
with $\frac{1}{4}<\alpha\leq\frac{1}{3}$, which corresponds to the hyperplane (\ref{eq34}). We claim that the $W_{\alpha}$ and $W^{'}_{\alpha}$ are nd-EWs. To prove this, we introduce the following state
\begin{equation}\label{eq39}
\rho=a_{1}O_{1}+a_{2}O_{2}+a_{3}O_{3}+a_{4}O_{4}
\end{equation}
in which $a_{i}\geq0, i=1,2,3,4$ and $a_{1}+a_{2}+a_{3}+a_{4}=1$. The PPT conditions for this state are $a_{1}a_{3}\geq a^{2}_{4}$ and $a_{2}\geq a_{4}$. The expectation values of $W_{\alpha}$ and $W^{'}_{\alpha}$ with respect to the $\rho$ are respectively
\begin{equation}\label{eq40}
Tr(W_{\alpha}\rho)=(a_{1}-a_{4})(1-\frac{1}{4\alpha})
\end{equation}
and
\begin{equation}\label{eq41}
Tr(W^{'}_{\alpha}\rho)=(a_{3}-a_{4})(1-\frac{1}{4\alpha})
\end{equation}
As a special case of the state (\ref{eq39}), we consider the following state
\begin{equation}\label{eq42}
\varrho=\frac{\beta}{13+\gamma}O_{1}+\frac{\gamma}{13+\gamma}O_{2}+\frac{10-\beta}{13+\gamma}O_{3}+\frac{3}{13+\gamma}O_{4}
\end{equation}
where $0\leq\beta\leq10$ and $\gamma$ is non-negative. It is PPT for $1\leq\beta\leq9$ and $\gamma\geq3$ and not PPT for $\beta<1$, $\beta>9$ or $\gamma<3$. Taking the expectation values of $W_{\alpha}$ and $W^{'}_{\alpha}$ with $\varrho$ shows that $\varrho$ is PPT entangled state for $1\leq\beta<3$, $7<\beta\leq9$ and $\gamma\geq3$. For $0\leq\beta<1$ and $9<\beta\leq10$, it is free entangled and for $3\leq\beta\leq7$ and $\gamma\geq3$ we can not say anything about its separability. The EW corresponding to the hyperplane (\ref{eq35}) is
\begin{equation}\label{eq43}
W^{''}_{\alpha}=I_{4}\otimes I_{4}-\frac{1}{\alpha}O_{2}-4(O_{1}+O_{3})-(2-\frac{1}{4\alpha})O_{4},
\end{equation}
whose expectation value with the $\varrho$ shows that it is free entangled for $\gamma<3$ and does not detect it in the PPT regime. Finally, the EW corresponding to the hyper-plane (\ref{eq36}) is
\begin{equation}\label{eq44}
W^{'''}_{\alpha}=I_{4}\otimes I_{4}-4(O_{1}+O_{2}+O_{3})-\frac{1}{\alpha}O_{4}
\end{equation}
For $\alpha\in(-\infty , 0)\bigcup[1 , \infty)$ it is always a positive operator and for $\alpha=0$, we have $W^{'''}_{0}=O_{4}$ which is also a positive operator. To give a physical meaning for the EWs  $W_{\alpha}$ and $W_{\alpha'}$ in detecting the entanglement of a state experimentally, from a measurement point of view, as in the previous section, these EWs should be decomposed into a sum of locally measurable operators. There exist following decompositions for $W_{\alpha}$ and $W_{\alpha'}$ as
$$W_{\alpha}=\frac{24\alpha-3}{64\alpha}I_{4}\otimes I_{4}+\frac{8\alpha-1}{32\alpha}(\lambda_{2}\otimes\lambda_{2}-\lambda_{1}\otimes\lambda_{1}+\lambda_{5}\otimes\lambda_{5}-\lambda_{4}\otimes\lambda_{4}+\lambda_{7}\otimes\lambda_{7}-\lambda_{6}\otimes\lambda_{6}$$
$$+\lambda_{10}\otimes\lambda_{10}-\lambda_{9}\otimes\lambda_{9}+\lambda_{12}\otimes\lambda_{12}-\lambda_{11}\otimes\lambda_{11}+\lambda_{14}\otimes\lambda_{14}-\lambda_{13}\otimes\lambda_{13})+\frac{3}{32\alpha}\lambda_{3}\otimes\lambda_{3}$$
$$+\frac{16\alpha+5}{96\alpha}\lambda_{8}\otimes\lambda_{8}+\frac{8\alpha+7}{96\alpha}\lambda_{15}\otimes\lambda_{15}+\frac{\sqrt{3}(4\alpha-1)}{16\alpha}(\lambda_{3}\otimes\lambda_{8}-\lambda_{8}\otimes\lambda_{3})$$
\be\label{eq29}+\frac{\sqrt{2}(4\alpha-1)}{12\alpha}(\lambda_{8}\otimes\lambda_{15}-\lambda_{15}\otimes\lambda_{8})+\frac{\sqrt{6}(1-4\alpha)}{24\alpha}(\lambda_{15}\otimes\lambda_{3})\ee and
$$W_{\alpha'}=\frac{24\alpha'-3}{64\alpha'}I_{4}\otimes I_{4}+\frac{8\alpha'-1}{32\alpha'}(\lambda_{2}\otimes\lambda_{2}-\lambda_{1}\otimes\lambda_{1}+\lambda_{5}\otimes\lambda_{5}-\lambda_{4}\otimes\lambda_{4}+\lambda_{7}\otimes\lambda_{7}-\lambda_{6}\otimes\lambda_{6}$$
$$+\lambda_{10}\otimes\lambda_{10}-\lambda_{9}\otimes\lambda_{9}+\lambda_{12}\otimes\lambda_{12}-\lambda_{11}\otimes\lambda_{11}+\lambda_{14}\otimes\lambda_{14}-\lambda_{13}\otimes\lambda_{13})+\frac{3}{32\alpha'}\lambda_{3}\otimes\lambda_{3}$$
$$+\frac{16\alpha'+5}{96\alpha'}\lambda_{8}\otimes\lambda_{8}+\frac{8\alpha'+7}{96\alpha'}\lambda_{15}\otimes\lambda_{15}+\frac{\sqrt{3}(4\alpha'-1)}{48\alpha'}(3\lambda_{8}\otimes\lambda_{3}-\lambda_{3}\otimes\lambda_{8})$$
\be\label{eq29}+\frac{\sqrt{2}(4\alpha'-1)}{24\alpha'}(2\lambda_{15}\otimes\lambda_{8}-\lambda_{8}\otimes\lambda_{15})+\frac{\sqrt{6}(1-4\alpha')}{24\alpha'}(\lambda_{3}\otimes\lambda_{15}),\ee
where $\lambda_{i}$s are the basis for the $su(4)$ Lie algebra
\cite{su(n)}. Therefore to measure the EWs $W_{\alpha}$ and
$W_{\alpha'}$, Alice and Bob need twenty local measurement
settings. At the end of the paper, it should be noted that in
order to characterize entanglement properties of other PPT states
in $3\otimes3$ and $4\otimes4$ systems, and for higher dimensional
bipartite one, several efforts have been made elsewhere(see  e.g.
\cite{heinz1,heinz2,heinz3}).
\section{Conclusions}
We have constructed new sets of EWs for the $3\otimes3$ systems analytically and for the $4\otimes4$ ones numerically. These witnesses are able to detect entanglement of some states in PPT regime. It has been shown that the existence of such EWs directly depends on the existence of a plane or a hyperplane as an exact boundary of the feasible region for the $3\otimes3$ and $4\otimes4$ systems respectively. The intersections of the plane or hyperplane with other boundaries of the feasible region are line segments or restricted planes respectively. As described, the rotation of the plane or hyperplane around the line segments or restricted planes in the allowed range leaves them in contact with, without intersecting, the related feasible region at the mentioned line segments or restricted planes. This situation, in fact, is due to the non-differentiability of the boundaries of the feasible region at the intersections. In this way, we obtain an infinite number of EWs. The approach can be generalized to the $n\otimes n$ systems. To this aim, one can choose the operators $O_{1}=\frac{1}{n}\sum_{i=0}^{n-1}(I_{n}\otimes S)|ii\rangle\langle ii|(I_{n}\otimes S^{\dagger})$, $O_{2}=\frac{1}{n}\sum_{i=0}^{n-1}(I_{n}\otimes S^{2})|ii\rangle\langle ii|(I_{n}\otimes S^{2\dagger})$,..., $O_{n-1}=\frac{1}{n}\sum_{i=0}^{n-1}(I_{n}\otimes S^{(n-1)})|ii\rangle\langle ii|(I_{n}\otimes S^{(n-1)\dagger})$ and $O_{n}=|\psi\langle\psi|$  as a basis for constructing EWs where $S$ is the shift operator. Therefore, it is expected that there exists a hyperplane as an exact boundary of the feasible region of the form $n(p_{1}+p_{2}+p_{3}+\cdots+p_{n-1})+p_{n}=1$. The existence of such hyperplane motivates us to construct similar witnesses as constructed for the $3\otimes3$ and $4\otimes4$. If such witnesses are constructed, they will detect the entanglement of the generalized Horodecki state, i.e. $\rho=\sum_{i=1}^{n}a_{i}O_{i}$, in the PPT regime.

\begin{table}[h]
\renewcommand{\arraystretch}{1}
\addtolength{\arraycolsep}{0pt}
$$
\small{
\begin{array}{|c|c|c|c|}\hline
  P_{1} & P_{2} & P_{3} & \mathrm{Product \ state}
  \\ \hline
  \frac{1}{3} & 0 & 0 &
 \tiny{\left(%
\begin{array}{c}
  1 \\
  0 \\
  0 \\
\end{array}
\right)}\otimes\tiny{\left(%
\begin{array}{c}
  0 \\
  1 \\
  0 \\
\end{array}
\right)}\\
  0 & \frac{1}{3} & 0 & \tiny{\left(%
\begin{array}{c}
  1 \\
  0 \\
  0 \\
\end{array}
\right)}\otimes\tiny{\left(%
\begin{array}{c}
  0 \\
  0 \\
  1 \\
\end{array}
\right)}\\
  0 & 0 & \frac{1}{3}& \tiny{\left(%
\begin{array}{c}
  1 \\
  0 \\
  0 \\
\end{array}
\right)}\otimes\tiny{\left(%
\begin{array}{c}
  1 \\
  0 \\
  0 \\
\end{array}
\right)}\\
  \frac{1}{9} & \frac{1}{9} & \frac{1}{3} &\tiny{\frac{1}{\sqrt{3}}\left(%
\begin{array}{c}
  1 \\
  1 \\
  1 \\
\end{array}
\right)}\otimes\tiny{\frac{1}{\sqrt{3}}\left(%
\begin{array}{c}
  1 \\
  1 \\
  1 \\
\end{array}
\right)}\\
  \frac{1}{48} & \frac{3}{16} & \frac{1}{4} &\tiny{\frac{1}{2}\left(%
\begin{array}{c}
  0 \\
  \sqrt{3} \\
  1\\
\end{array}
\right)}\otimes\tiny{\frac{1}{2}\left(%
\begin{array}{c}
  0 \\
  1 \\
  \sqrt{3} \\
\end{array}
\right)}\\
  \frac{1}{192} & \frac{49}{192} & \frac{7}{48} &\tiny{\frac{1}{4}\left(%
\begin{array}{c}
  0 \\
  \sqrt{2} \\
  \sqrt{14}\\
\end{array}
\right)}\otimes\tiny{\frac{1}{4}\left(%
\begin{array}{c}
  0 \\
  \sqrt{14} \\
  \sqrt{2} \\
\end{array}
\right)}\\
  \frac{3}{64} & \frac{25}{192} & \frac{5}{16} &\tiny{\frac{1}{4}\left(%
\begin{array}{c}
  0 \\
  \sqrt{6} \\
  \sqrt{10}\\
\end{array}
\right)}\otimes\tiny{\frac{1}{4}\left(%
\begin{array}{c}
  0 \\
  \sqrt{10} \\
  \sqrt{6} \\
\end{array}
\right)}\\
 \frac{1}{12} & \frac{1}{8} & \frac{1}{3} & \mathrm{no\ product}\\
  \frac{3}{16} & \frac{1}{48} & \frac{1}{4} & \tiny{\frac{1}{2}\left(%
\begin{array}{c}
  0 \\
  1 \\
  \sqrt{3}\\
\end{array}
\right)}\otimes\tiny{\frac{1}{2}\left(%
\begin{array}{c}
  0 \\
  \sqrt{3} \\
  1 \\
\end{array}
\right)}\\
  \frac{49}{192} & \frac{1}{192} & \frac{7}{48} &\tiny{\frac{1}{4}\left(%
\begin{array}{c}
  0 \\
  \sqrt{14} \\
  \sqrt{2}\\
\end{array}
\right)}\otimes\tiny{\frac{1}{4}\left(%
\begin{array}{c}
  0\\
  \sqrt{2} \\
  \sqrt{14} \\
\end{array}
\right)}\\
      \frac{25}{192} & \frac{3}{64} & \frac{5}{16} &\tiny{\frac{1}{4}\left(%
\begin{array}{c}
  0 \\
  \sqrt{10} \\
  \sqrt{6}\\
\end{array}
\right)}\otimes\tiny{\frac{1}{4}\left(%
\begin{array}{c}
  0\\
  \sqrt{6} \\
  \sqrt{10} \\
\end{array}
\right)}\\
      \frac{1}{8} & \frac{1}{12} & \frac{1}{3}& \mathrm{no\ product}\\
       \hline
\end{array}}
$$
\caption{\small{The product states and coordinates of vertices
for $3\otimes3$ EWs.}}
\renewcommand{\arraystretch}{1}
\addtolength{\arraycolsep}{-3pt}
\end{table}

\end{document}